\documentclass[8.5pt,twoside,twocolumn]{article}
\usepackage[super,sort&compress,comma]{natbib} 
\usepackage{mhchem}
\usepackage{times,mathptmx}
\usepackage{sectsty}
\usepackage{balance} 

\usepackage{graphicx} 
\usepackage{lastpage}
\usepackage[format=plain,justification=raggedright,singlelinecheck=false,font=small,labelfont=bf,labelsep=space]{caption} 
\usepackage{fancyhdr}
\begin{document}

\thispagestyle{plain}
\fancypagestyle{plain}{
\renewcommand{\headrulewidth}{1pt}}
\renewcommand{\thefootnote}{\fnsymbol{footnote}}
\renewcommand\footnoterule{\vspace*{1pt}%
\hrule width 3.4in height 0.4pt \vspace*{5pt}} 
\setcounter{secnumdepth}{5}

\makeatletter 
\def\subsubsection{\@startsection{subsubsection}{3}{10pt}{-1.25ex plus -1ex minus -.1ex}{0ex plus 0ex}{\normalsize\bf}} 
\def\paragraph{\@startsection{paragraph}{4}{10pt}{-1.25ex plus -1ex minus -.1ex}{0ex plus 0ex}{\normalsize\textit}} 
\renewcommand\@biblabel[1]{#1}            
\renewcommand\@makefntext[1]%
{\noindent\makebox[0pt][r]{\@thefnmark\,}#1}
\makeatother 
\renewcommand{\figurename}{\small{Fig.}~}
\sectionfont{\large}
\subsectionfont{\normalsize} 

\fancyfoot{}
\fancyhead{}
\renewcommand{\headrulewidth}{1pt} 
\renewcommand{\footrulewidth}{1pt}

\newcommand{\unit}[1]{\ensuremath{\, \mathrm{#1}}}

\setlength{\arrayrulewidth}{1pt}
\setlength{\columnsep}{6.5mm}
\setlength\bibsep{1pt}

\twocolumn[
  \begin{@twocolumnfalse}
\noindent\LARGE{\textbf{Spinning nanorods -- active optical manipulation of semiconductor nanorods using polarised light}}
\vspace{0.6cm}

\noindent\large{\textbf{C. Robin Head,\textit{$^{a}$} Elena Kammann,\textit{$^{a}$} Marco Zanella,\textit{$^{b}$} Liberato Manna,\textit{$^{b,c}$} and
Pavlos G. Lagoudakis\textit{$^{a\ast}$}}}\vspace{0.5cm}



\noindent \normalsize{In this letter we show how a single beam optical trap offers the means for three-dimensional manipulation of semiconductor nanorods in solution. Furthermore rotation of the direction of the electric field provides control over the orientation of the nanorods, which is shown by polarisation analysis of two photon induced fluorescence. Statistics over tens of trapped agglomerates reveal a correlation between the measured degree of polarisation, the trap stiffness and the intensity of the emitted light, confirming that we are approaching the single particle limit.}
\vspace{0.5cm}
 \end{@twocolumnfalse}
  ]

\section{Introduction}


\footnotetext{\textit{$^{a}$School of Physics and Astronomy, University of Southampton, Southampton SO17 1BJ, United Kingdom}}
\footnotetext{\textit{$^{b}$Instituto Italiano di Tecnologia, via Morego 30, 16163 Genova, Italy                                                                                                   }}
\footnotetext{\textit{$^{c}$3 Kavli Institute of Nanoscience, Delft University of Technology, Lorentzweg 1, 2628 CJ Delft, The Netherlands                                                            }}
\footnotetext{\textit{$^\ast$ correspondence to:} Pavlos.Lagoudakis@soton.ac.uk                                                           }
Core-shell semiconductor nanorods are engineered nanoparticles with unusual fluorescence properties.\cite{Muller:2005ed,Carbone:2007ug} Their asymmetry imposes a defined linear polarisation on the emitted photons\cite{Chen:2001ky,Hu:2001hs} and therefore they are promising candidates for polarised single photon emitters for quantum computing and a possible application as electric field optical nanosensors was demonstrated.\cite{Muller:2005ed}  
Optical tweezing of quantum dots and rods has recently generated a lot of interest\cite{Jauffred:2008wa,Jauffred:2010cx,Ramachandran:2005vp} as it has a vast range of applications in biology and material science. Optical traps can be used for high accuracy particle delivery and printing.\cite{Yu:2004kj} In conjunction with polarisation induced alignment of colloidal quantum rods, they can be a powerful tool for nanomanipulation, nanofabrication and optical nanosensing. Moreover it allows optical experiments on single nanorods in solution, a technique which has been demonstrated for gold nanoparticles.\cite{Prikulis:2004kk} 
Previous works have shown that angular momentum of light can cause spinning of birefringent\cite{Friese:640243} or asymmetric particles.\cite{Tong:2009jz} Furthermore linear polarisation can align particles in optical traps\cite{Jauffred:2008wa,Yu:2004kj,SelhuberUnkel:2008fw,Pelton:2006wv} and rotation of the linear polarisation causes trapped particles to spin at the rotation frequency.\cite{Tong:2009jz,Jones:2009jg,Bonin:2002tv} 
The most widely used tool for detection and quantification of micro and nanoparticles in single beam optical traps is a quadrant photo diode.\cite{Oddershede:466548} Forward scattered light which contains information about the position of the particle allows for extraction of the trap stiffness. The trap stiffness in different directions has previously been used to infer the alignment of the trapped particle\cite{Jauffred:2008wa,SelhuberUnkel:2008fw} when it is too small to be imaged by optical means. Here we exploit an alternative detection mechanism: the polarised emission along the long axis of semiconductor nanorods (based on core/shell CdSe/CdS nanocrystals) is utilized to show the alignment of the particle in the trap. The degree of the linear polarisation (($DLP=(I_{max}-I_{min})/(I_{max}+I_{min})$), where $I_{max/min}$ denotes the maximum and minimum intensity measured upon rotation of the polarisation analyser) provides insight into the nature of the trapped particle. A single particle will exhibit a high degree of linear polarisation, comparable to a single particle on a glass slide. Agglomerates of a few particles will show a reduced $DLP$, which is however still greater than zero. 

\section{Methods}
\begin{figure}[htb]
\centering
  \includegraphics{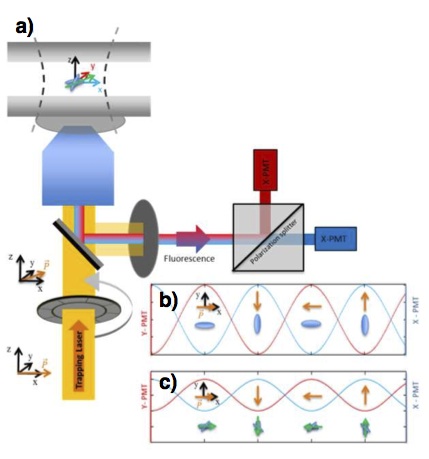}
  \caption{Schematic of the experiment: a The nanorods are trapped by a NIR CW laser that has its polarisation in the x-y plane rotated by a half-wave plate. Asymmetric nanoparticles align with the linear polarisation of the trapping laser. A NIR cold mirror separates the quantum dot emission, a shortpass filter blocks scattered NIR light and the polarisation of the emission is analysed by two balanced Photomultipliers (PMTs) that record the x- and y-polarised light. b Schematic of the signal from the aligned nanorod. Single nanorods emit light, which is polarised along their long axis. Upon rotation of the input polarisation the particle follows the orientation of the polarisation. This can be observed by a sinusoidal modulation in the signal on the PMTs which anti-correlate with respect to each other. The amplitude of this modulation is close to the amplitude measured for a single particle on a glass slide. c Same as in b, but for an agglomerate of a few nanoparticles. Here the observed degree of polarisation is reduced due to the different orientations of the individual nanorods in the agglomerate.}
  \label{fig1}
\end{figure}

In Fig. \ref{fig1}a the optical trapping setup is shown schematically. A $1085 \unit{nm}$ $4\unit{W}$ N-light fibre laser is focused through the oil immersion objective ($NA=1.4$, Immersion medium: Cargille Type LDF immersion oil with a high refractive index of 1.5239). The sample solutions were held in a homemade liquid cell consisting of two cover slides and spacers. Particles in the optical trap were selectively excited through two-photon absorption of the trapping laser.\cite{Jauffred:2010cx} This non-resonant excitation mechanism ensures that the polarisation is not inherited from the pump laser. The emission was collected through the trapping objective and was spectrally separated from the trapping beam using an infrared cold mirror. Then the fluorescence was split into horizontal and vertical polarisation and simultaneously detected by two photomultiplier tubes (PMTs). The polarised trapping laser causes the particle to align in the optical trap along the linear polarisation. The direction of the linear polarisation was controlled by a remote controlled half-wave plate, that could be rotated at up to $80 \unit{Hz}$ (corresponding to a polarisation rotation frequency of $f_{HV}=320\unit{Hz}$). If the particle rotates the fluorescence signal on the PMTs will oscillate with the same frequency and a phase shift of $\pi$ between the two PMTs (schematically shown for a single nanorod (Fig. \ref{fig1}b) and a small agglomerate with a reduced degree of polarisation (Fig. \ref{fig1}c)). Additionally the particle solution could be imaged by fluorescence excitation with a blue laser diode that was focused to a large spot from the top. 
The trap stiffness was estimated by shaking the sample-chamber laterally with respect to the optical trap with an amplitude of $A=20\unit{\mu m}$ using a piezo-controlled stage. The movement of the stage was controlled with a signal generator that applied a continuous triangular function with a variable frequency. The speed of the solution with respect to the optical trap is given by $v = f \times A \times 2$, where $f$ is the triangular wave oscillation frequency.  The upper frequency limit at which the trapped particle left the trap, ergo when the drag force of the solution exceeded the lateral trapping force, is used to estimate the lateral trapping force using Stokes' law $F_d = 6\pi\mu Rv$ \cite{Oddershede:466548}, where $F_d$ is is the drag force exerted, $\mu$ the viscosity of the solution, $R$ the radius of the particle. 
We use CdSe/CdS core shell nanorods with a cylindrical geometry capped with two half spheres, with a high aspect ratio of 10 ($33.5  \pm 3.5 \unit{nm}/ 2.5  \pm 0.7\unit{nm}$). The rods were synthesized as described by Carbone and co-workers\cite{Carbone:2007ug} and then made water-soluble according to a published procedure which involved wrapping of the particles in an amphiphilic polymer shell.\cite{DiCorato:2008cb}
\section{Results and Discussion}
\begin{figure*}[htb]
\centering
  \includegraphics[width=0.9\textwidth]{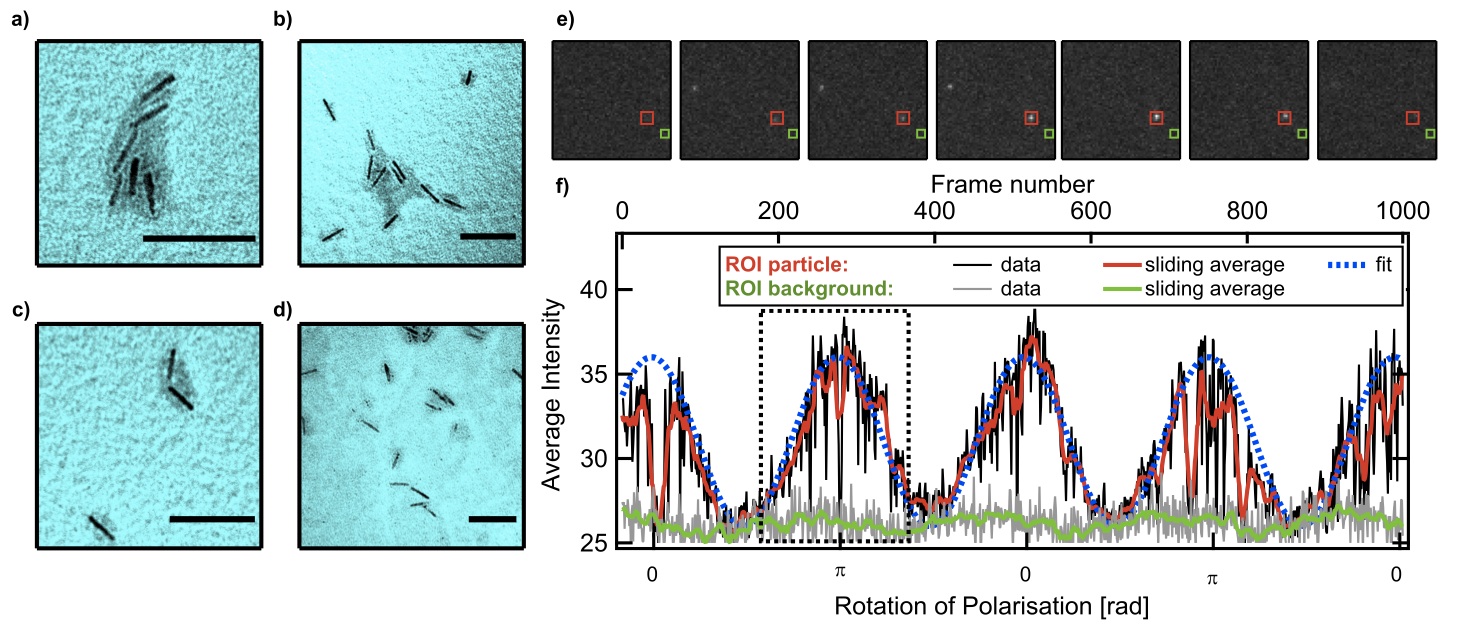}
  \caption{Nanorods -- from single particles to small agglomerates: a-d Transmission electron micrographs of the nanorods. Ligand material causes nanoparticles to form small agglomerates of for example 6 (a), 8 (b) or 2 (c) nanorods. Black bars are $100 \unit{nm}$ long. e Selected frames showing the measured regions over a $\pi$ turn of the polarisation. (f) polarisation resolved emission of a single nanorod on a glass slide. The red trace is a sliding average over 10 points for the average intensity of the region of interest (ROI) containing a Nanorod (marked in red in e), whereas the green trace shows the average intensity of a ROI next to the Nanorod (green square in e). The blue trace is a $cos^2$ fit, which reveals a polarisation degree of nearly $100\%$. }
  \label{fig2}
\end{figure*}

When imaging the nanoparticles in solution we found a certain inhomogeneity in particle brightness, which leads to the conclusion that some of these particles are in fact agglomerates of several nanorods. These small agglomerates were equally observed in transmission electron micrographs (TEM), as shown in Fig. \ref{fig2}a-d. For sufficiently small agglomerates with high anisotropy (Fig. \ref{fig2}a,b and c) it is still expected that the particle aligns in the optical trap and that the emission is polarised, albeit to a smaller degree than for a single particle. 
The polarisation properties of single nanorods were investigated by spin coating a low density solution of nanoparticles onto a coverslip and imaging the polarisation resolved fluorescence on an electron-multiplication CCD. The polarisation analyser was rotated at $0.1\unit{Hz}$. The average intensity measured, for a region of interest containing a single nanorod, at each frame, is displayed in Fig. \ref{fig2}f. A squared cosine fit to the data allows for extraction of the linear polarisation degree of nearly $100\%$. This quantity is crucial for comparison with the polarisation degrees obtained from particles in the optical trap. Additionally the data exhibits blinking, which confirms that the emission is indeed originating from a single nanorod.
\begin{figure}[htb]
\centering
  \includegraphics[width=0.4\textwidth]{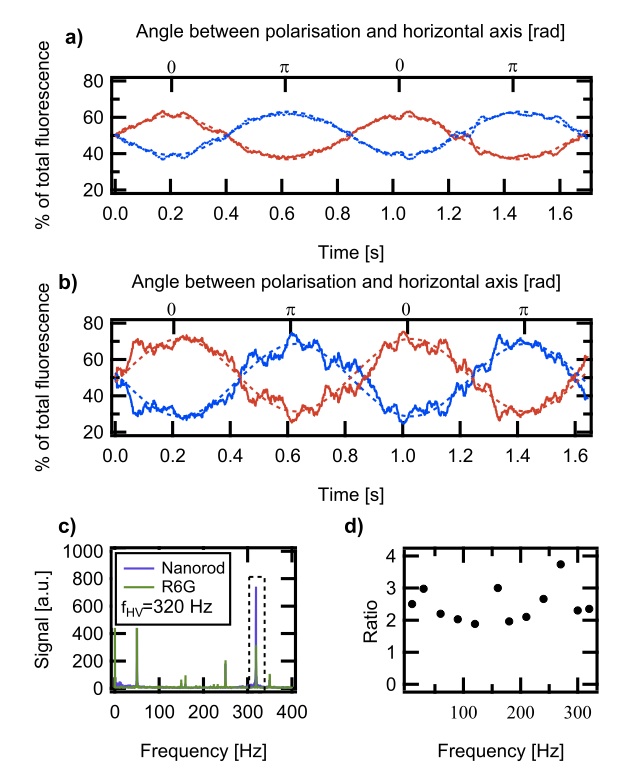}
  \caption{Polarisation spectroscopy of spinning nanorods: a,b Signal on the two PMTs from a rotating nanorod as a function of time/ degree of polarisation with respect to the horizontal axis at $1.25\unit{Hz}$ with a sliding average over 200 points (solid) and a sinusoidal fit (dashed). c Normalized Fourier spectra of the signal from the PMT at a rotation speed of $320\unit{Hz}$ for the nanorods (purple) and rhodamine 6G (green). The peak at $50 \unit{Hz}$ originates from the mains noise and is greater than the signal itself requiring the FFT data extraction.
d The ratio of the peaks at rotation frequency, for different rotation frequencies. This quantity remains strictly greater than one, confirming that the trapped particles follow the linear polarisation up to $320\unit{Hz}$. 
 }
  \label{fig3}
\end{figure}
Fig. \ref{fig3}a,b display intensity traces of the two balanced photomultiplier tubes, recording the horizontal (red) and vertical (blue) polarisation of a trapped particle as a function of time (bottom) and angle of polarisation with respect to the horizontal axis (top) with a sliding average over 200 points ($0.02\unit{s}$) (the polarisation was rotated at $1.25 \unit{Hz}$). When the trap was focused onto a dense solution of rhodamine 6G instead, a modulation of the signal was equally observed, corresponding to the rotation frequency $f_{HV}$. However in this case the signal from both PMTs oscillated with the same phase, which indicates a small polarisation dependent intensity fluctuation of the trapping beam. A rotation frequency-dependence was then performed in order to determine if the aligned particles can follow the rotation of the polarisation up to a frequency $f_{HV}=320\unit{Hz}$, which is the maximum frequency achievable with our apparatus. The polarisation of the trapping beam was rotated at different frequencies ($f_{HV}$). As the signal from the rotating particles was contaminated by the mains noise frequency (50Hz) the normalized signal was analysed with a Fast-Fourier-Transformation (FFT) shown in Fig. \ref{fig3}c. We compared the heights of the FFT peak at the rotation frequency with the normalized Fourier-spectrum of the results obtained for rhodamine 6G, to ensure that the signal modulation is not caused by an intensity modulation of the trapping beam. In Fig. \ref{fig3}d the ratio of the amplitude of $f_{HV}$ of the rotated nanorod ($A(f_{HV},NR)$) and the rhodamine ($A(f_{HV},RHO)$) is plotted as a function of rotation frequency. The ratio remained strictly greater than 1, confirming that the trapped particles can follow the rotation frequency up to $f_{HV}=320\unit{Hz}$ and that there was no significant dependence on the rotation frequency in the measured range. 
From the intensity traces of the trapped particles vs time (as displayed in Fig. \ref{fig3}a), it is possible to extract the degree of linear polarisation ($DLP$). Fig. \ref{fig4} displays a scatter plot of the $DLP$ for 40 trapped particles vs the absolute emission intensity (Fig. \ref{fig4}a) and the escape frequency (Fig. \ref{fig4}b). We found that the $DLP$ has a tendency to be lower for brighter particles and higher escape frequencies. This observation confirms that smaller agglomerates have a higher anisotropy and therefore exhibit a large $DLP$. We measured $DLP$s of up to $51.5\%$ for the smallest of particles in the optical trap. Even though this is only half of the $DLP$ of the single nanorod on a glass slide one must consider that the particles cannot be completely immobilised in the optical trap.\cite{Jauffred:2008wa} The polarisation induced alignment competes with the Brownian motion of the trapped particle. These thermal fluctuations cause a reduction of the $DLP$.
At last we estimated the size of particles of the trapped agglomerate, by introducing an effective Radius $R_{eff}$  that corresponds to a spherical particle with the same polarisability as the randomly shaped object in the optical trap. The polarisability of the object can then be approximated as the polarisability of a spherical particle with the radius $R_{eff}$
\begin{equation}
\alpha=4 \pi n_2^2  \epsilon_0  R_{eff}^3  \frac{m^2-1}{m^2+2}
\end{equation}
, where $n_2$ is the refractive index of the solution, $\epsilon_0$ the dielectric constant, $n_1$ the refractive index of the particle and $m=\tfrac{n_1}{n_2}$. We calculate the lateral trapping strength using the Rayleigh scattering model:\cite{Harada:1996va} 
\begin{equation}
F_{grad} (x,z)= \frac{-\alpha}{2c\epsilon_0}   \frac{I(x,z)}{1+2z^2 } \frac{4 x}{w_0} 
\end{equation}
Where $\alpha$ is the polarisability, $w_0$ the beam diameter at the waist and c the speed of light. The trapping laser intensity is given by
\begin{equation} 
I (x,z)=\frac{2 P}{\pi w_0^2 }  \frac{1}{1+2z^2 } e^{-\tfrac{(2x^2)}{(1+2z^2 )} }
\end{equation}
, where $P$ is the laser power. We set to the equilibrium position of the particle $z=\tfrac{2 \pi}{\lambda}\tfrac{w_0^2}{2\sqrt{3}}$, where gradient and scattering force are of equal size and opposite direction. Known parameters from the experiment are $P=0.55\unit{W}$, $n_1 = 2.37$, $n_2 = 1.332$, $\lambda = 1085 \unit{nm}$ and $w_0=1 \unit{\mu m}$. Using the Stokes' law we can estimate the escape frequency as a function of particle size. The obtained curve is plotted over the data that correlates the particle intensity and the escape frequency (Fig. \ref{fig4}c). The escape frequencies for a single spherical particle of the same volume as the nanorods and an agglomerate of 10 particles are indicated in the figure. This calculation however neglects the ligand that forms the agglomerate, as visible on the TEM pictures (Fig. \ref{fig2}a-d). The ligand material contributes to the volume and hence to the trapping strength, ultimately increasing the escape frequency, leading to an overestimation of the number of particle within an agglomerate. Thus our calculated smallest agglomerate formed out of 10 particles is actually likely to be smaller. Moreover it has previously been demonstrated, that quantum dots can be trapped with much less power, than theoretically predicted.\cite{Jauffred:2008wa}
\begin{figure}[htb]
\centering
  \includegraphics[width=0.4\textwidth]{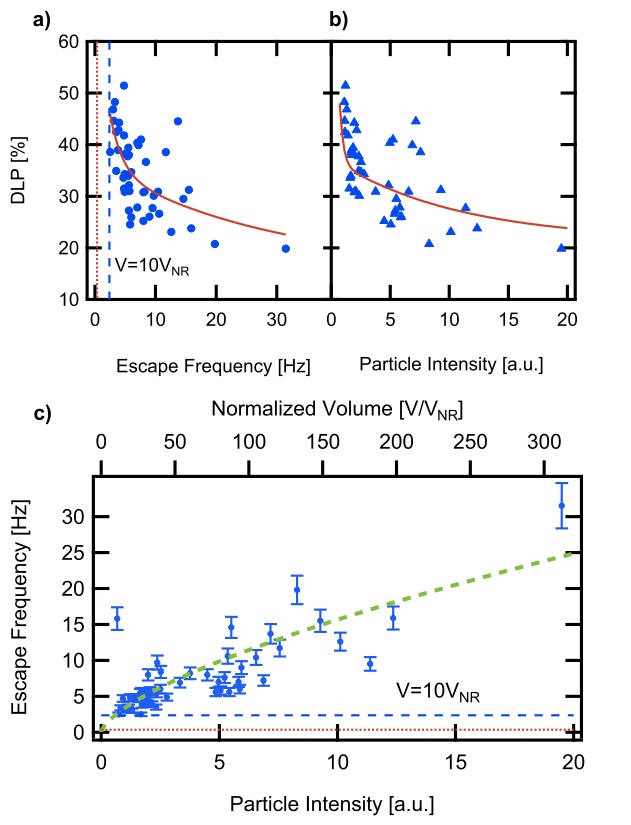}
  \caption{Statistics of particles in the optical trap: Scatterplots of $DLP$ vs intensity emitted by the particle in the optical trap (a) and the escape frequency (b) measured for 48 separately trapped objects. A higher degree of polarisation is observed for particles with lower intensity and smaller escape frequencies. c Scatterplot of escape frequency vs particle intensity. The green dashed line is the calculated dependency of the escape frequency on the volume of the particles overlain with the data. The vertical (horizontal) lines in a (c) indicate the escape frequencies expected for a spherical particle of the same volume as one (red dotted) and ten (dashed blue) of our nanorods.}
  \label{fig4}
\end{figure}
\section{Conclusions}
In conclusion we have optically trapped and aligned agglomerates of CdSe/CdS nanorods, which was demonstrated by polarisation resolved fluorescence detection. Statistics over tens of particles show a correlation between intensity, escape frequency and the degree of linear polarisation. Moreover it has been shown that the agglomerates rotate up to rotation frequencies of 320 Hz. Our observations pave way to nanomotors, controlled assembly of nanorods to functional materials and specific positioning of optical nanosensors with control over the alignment of the nanoparticle.

\section{Acknowledgements}
We thank Lene Oddershede and Liselotte Jauffred for discussions and advice and Giammarino Pugliese for help with the procedure of nanorod transfer in water. We acknowledge funding from FP7 ITN Spinoptronics and the Network of Excellence N4E.




\footnotesize{
\bibliography{trapping} 
\bibliographystyle{rsc} 
}

\end{document}